\documentclass[aps,prc,twoside,twocolumn,floatfix,superscriptaddress,showpacs]
{revtex4}
\usepackage{amssymb}
\usepackage{amsmath}
\usepackage{graphicx}
\usepackage{lscape}
\usepackage{color}
\def\rpi{$\pi^-/\pi^+$~}

\def\esym{$E_{\rm sym}(\rho)$~}

\def\rnz{$\left<N/Z\right>$~}

\def\pmi{$\pi^-$~}

\def\agev{GeV/nucleon~}

\begin{document}

\title{A systematic study of the $\pi^-/\pi^+$ ratio in heavy-ion collisions with the same neutron/proton ratio but different masses}

\author{Ming Zhang}\affiliation{Department of Physics, Tsinghua University, Beijing 100084,  China}
\author{Zhi-Gang Xiao}\email {xiaozg@tsinghua.edu.cn}\affiliation{Department of Physics, Tsinghua University, Beijing 100084,  China}
\author{Bao-An Li}\affiliation{Department of Physics and Astronomy, Texas A\& M University-Commerce, Commerce, Texas 75429-3011, USA}
\author{Lie-Wen Chen}\affiliation{Department of Physics, Shanghai Jiao Tong University, Shanghai 200240, China}
\author{Gao-Chan Yong}\affiliation{Department of Physics and Astronomy, Texas A\& M University-Commerce, Commerce, Texas 75429-3011, USA}
\affiliation{Institute of Modern Physics, Chinese Academy of
Science, Lanzhou 730000, China}
\author{ Sheng-Jiang Zhu}\affiliation{Department of Physics, Tsinghua University, Beijing 100084,  China}

\date{\today}

\begin{abstract}
A systematic study of the $\pi^-/\pi^+$ ratio in heavy-ion
collisions with the same neutron/proton ratio but different masses
can help single out effects of the nuclear mean field on pion
production. Based on simulations using the IBUU04 transport model,
it is found that the $\pi^-$/$\pi^+$ ratio in head-on collisions of
$^{48}$Ca+$^{48}$Ca, $^{124}$Sn +$^{124}$Sn and
$^{197}$Au+$^{197}$Au at beam energies from 0.25  to 0.6 \agev
increases with increasing the system size or decreasing the beam
energies. A comprehensive analysis of the dynamical isospin
fractionation and the \rpi ratio as well as their time evolution and
spatial distributions demonstrates clearly that the \rpi ratio is an
effective probe of the high-density behavior of the nuclear symmetry
energy.

\end{abstract}
\pacs{25.70.-z, 25.60.-t, 25.80.Ls, 24.10.Lx}

\maketitle

The high-density (HD) behavior of the nuclear symmetry energy \esym
has long been regarded as among the most uncertain properties of
dense neutron-rich nuclear matter \cite{KUT94,BAL02,KUB03,BAL08}. It
is very essential for understanding not only many fundamental
astrophysical phenomena \cite{KSU95,JML04,AWS05} but also novel
features of high energy heavy ion reactions especially those induced
by rare isotopes \cite{BAL08,BAL01,BAL98,PDR02,VBM05}. The
$\pi^-/\pi^+$ ratio in heavy-ion collisions has been known as a
particularly sensitive probe of the HD behavior of the \esym
\cite{BAL02}. Based on the IBUU04 transport model \cite{IBUU04}
analysis of the $\pi^-/\pi^+$ data from the FOPI collaboration
\cite{WRE07}, circumstantial evidence suggesting a rather soft \esym
compared to the widely used APR ( Akmal-Pandharipande-Ravenhall)
prediction\cite{AAK98} was reported very recently \cite{ZGX08}. The
IBUU04 transport model used the single particle potential $U$
corresponding to a modified Gogny Momentum Dependent Interaction
(MDI) \cite{CBD03}, i.e., for a nucleon with momentum $\vec{p}$ and
isospin $\tau $,
\begin{eqnarray}
U(\rho,\delta ,\vec{p},\tau ) =A_{u}(x)\frac{\rho _{-\tau }}{\rho
_{0}} +A_{l}(x)\frac{\rho _{\tau }}{\rho _{0}} +B(\frac{\rho }{\rho
_{0}})^{\sigma }(1-x\delta ^{2}) \notag \\
-8\tau x\frac{B}{\sigma +1}\frac{\rho ^{\sigma -1}}{\rho
_{0}^{\sigma }}\delta \rho _{-\tau } +\frac{2C_{\tau ,\tau }}{\rho
_{0}}\int d^{3}p^{\prime }\frac{f_{\tau }( \vec{r},\vec{p}^{\prime
})}{1+(\vec{p}-\vec{p}^{\prime })^{2}/\Lambda ^{2}}
\notag \\
+\frac{2C_{\tau ,-\tau }}{\rho _{0}}\int d^{3}p^{\prime
}\frac{f_{-\tau }( \vec{r},\vec{p}^{\prime
})}{1+(\vec{p}-\vec{p}^{\prime })^{2}/\Lambda ^{2}}
\label{MDIU}
\end{eqnarray}

\begin{figure}[h]
\centering
\includegraphics[width=\columnwidth]{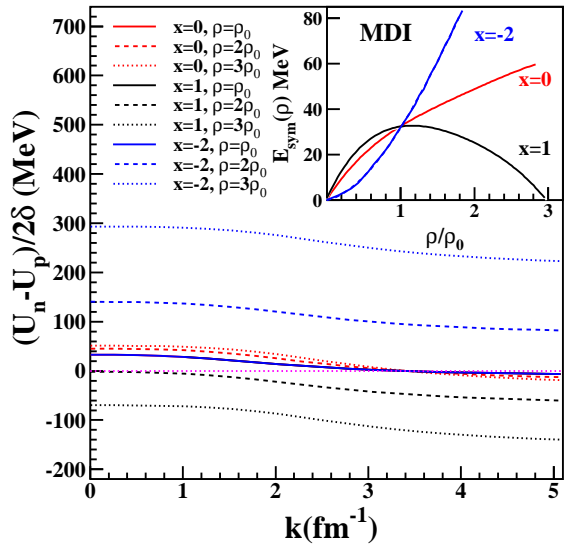}
\caption{(Color online) The symmetry potential as a function of
nucleon momentum. With $\rho=\rho_0$, the curves are identical for
x=-2, 0 and 1. The inset shows the density dependence of the
symmetry energy with x=-2, 0 and 1, respectively. }\label{fig1}
\end{figure}

In the above equation the isospin $\tau =1/2$ ($-1/2$) for neutrons
(protons). The coefficients $A_{u}(x)$ and $A_{l}(x)$ are
\cite{LWC05}
\begin{equation}\label{alau}
A_{u}(x)=-95.98-x\frac{2B}{\sigma
+1},~~~A_{l}(x)=-120.57+x\frac{2B}{\sigma +1}.
\end{equation}
The values of the parameters are $\sigma=4/3$, $B=106.35$ MeV,
$C_{\tau ,\tau }=-11.70$ MeV, $C_{\tau ,-\tau }=-103.40$ MeV and
$\Lambda=p_f^{0}$ which is the Fermi momentum of nuclear matter at
the saturation density $\rho_0$ \cite{CBD03}. For asymmetric
nuclear matter at zero temperature, the MDI symmetry energy can be
written as \cite{Xu09}
\begin{eqnarray}\label{esymmdi}
E_{sym}(\rho) = \frac{1}{2} \left(\frac{\partial^2 E}
{\partial \delta^2}\right)_{\delta=0} \notag\\
= \frac{8 \pi}{9 m h^3 \rho} p^5_f + \frac{\rho}{4 \rho_0} (A_l(x)-
A_u(x)) - \frac{B x}{\sigma + 1}
\left(\frac{\rho}{\rho_0}\right)^\sigma \notag\\
+ \frac{C_l}{9 \rho_0 \rho} \left(\frac{4 \pi}{h^3}\right)^2
\Lambda^2 \left[4 p^4_f - \Lambda^2 p^2_f \ln \frac{4 p^2_f
+ \Lambda^2}{\Lambda^2}\right] \notag\\
+ \frac{C_u}{9 \rho_0 \rho} \left(\frac{4 \pi}{h^3}\right)^2
\Lambda^2 \left[4 p^4_f - p^2_f (4 p^2_f + \Lambda^2) \ln \frac{4
p^2_f + \Lambda^2}{\Lambda^2}\right]
\end{eqnarray}
where $p_f=\hbar(3\pi^2\frac{\rho}{2})^{1/3}$ is the Fermi
momentum for symmetric nuclear matter at density $\rho$. We note
here that since the $A_l(x)-A_u(x)=-24.59+4Bx/(\sigma +1)$
according to Eq.~(\ref{alau}), the $E_{sym}(\rho)$ depends
linearly on the parameter $x$ at a given density except at
$\rho_0$ where the $E_{\rm sym}(\rho_0)$ is fixed at $30.54$ MeV.
Shown in the inset of Fig.\ 1 are examples of the \esym with x=1,
0 and -2, respectively.  The strengthes of the corresponding
isovector (symmetry) potential estimated from $(U_n-U_p)/2\delta$
are shown for 3 typical densities in the main frame of Fig.\
\ref{fig1}. At the normal density $\rho_0$, by design, the
symmetry potential is independent of x and is consistent with the
Lane potential extracted from the experimental data of
nucleon-nucleus scatterings and the (p,n) charge exchange
reactions \cite{BAL08}. It is necessary to emphasize that it is
the symmetry potential, not the symmetry energy, that enters as a
direct input in all transport model simulations. It is thus
important to point out the key characteristics of the symmetry
potentials. With x=0, the symmetry potential is weak but remains
mostly positive at all densities in the momentum range considered.
With x=-2 leading to the stiffer symmetry energy, the symmetry
potential is positive at all densities. With x=1 leading to the
softer symmetry energy, however, the symmetry potential is
negative at supra-saturation densities.

It was found that only the very soft \esym with $x=1$ can well
reproduce the FOPI data \cite{ZGX08}. The stiffer \esym with $x=0$
(which resembles very well the APR prediction up to about
$3.5\rho_0$ \cite{ZGX08}) under-predicts significantly the data as
in earlier IQMD \cite{IQMD} calculations \cite{WRE07} using an
\esym very similar to the APR prediction or the MDI \esym with
$x=0$. The interesting ramifications in both nuclear physics and
astrophysics of this finding about the HD behavior of the \esym
strongly calls for additional theoretical studies and experimental
tests. Since pions always undergo strong final state interactions,
the latter may distort the information carried by the
$\pi^-/\pi^+$ messenger about the HD \esym. In this work, based on
the IBUU04 transport model simulations of heavy-ion collisions
with the same neutron/proton ratio but different masses we further
investigate how reliable the $\pi^-/\pi^+$ ratio is in probing the
HD behavior of the \esym.

To understand the advantage of comparing systematically the
$\pi^-/\pi^+$ ratios in heavy-ion reactions with the same
neutron/proton ratio but different masses, we start by recalling
the first-chance nucleon-nucleon collision model for pion
production and the dynamical isospin fractionation mechanism for
enhancing the neutron/proton ratio of the HD region and thus the
$\pi^-/\pi^+$ ratio from there. Near the pion production threshold
of about 300 MeV, it is reasonable to assume that only first
chance inelastic nucleon-nucleon collisions produce pions. Since
the $\pi^-$ and $\pi^+$ are mostly produced from $nn\rightarrow
np+\pi^-$ and $pp\rightarrow np+\pi^+$, respectively, and the $pn$
collisions contribute equally to the production of both $\pi^-$
and $\pi^+$, the primordial $\pi^-/\pi^+$ ratio is expected to be
proportional to the neutron/proton ratio squared, i.e.,
$(N/Z)^2_{\rm dense}$, of the participant region. Indeed, more
detailed estimate based on the $\Delta (1232)$ isobar model
\cite{Sto86} predicts a primordial $\pi^-/\pi^+$ ratio of $R_{\rm
isob}\equiv (\pi^-/\pi^+)_{\rm res}\equiv
(5N^2+NZ)/(5Z^2+NZ)\approx (N/Z)^2_{\rm dense}$. Due to the
detailed balance, the reabsorption of $\pi^-$ and $\pi^+$ mainly
through the $\Delta$ resonances correlate proportionally to their
production, they are thus not expected to change significantly the
primordial $\pi^-/\pi^+$ ratio in the absence of possible
in-medium effects of the $\Delta$ resonances (see, however, refs.
\cite{LAR03} and \cite{JXU09}). The Coulomb force in the final
state is expected to affect the differential $\pi^-/\pi^+$ ratio
as a function of pion momentum but not the integrated
$\pi^-/\pi^+$ ratio. The final $\pi^-/\pi^+$ multiplicity ratio is
thus a direct measure of the isospin asymmetry $(N/Z)_{\rm dense}$
in the early stage of the reaction. While the latter is determined
by the HD \esym through the dynamical isospin fractionation
phenomenon \cite{Mul95,LiBA97b,Bar98,LiBA00,HSX00}, namely, the
participant region is more neutron-rich (poor) if the value of the
HD \esym there is lower (higher). In terms of the reaction
dynamics for pion production within transport models, the
$\pi^-/\pi^+$ ratio depends on the $(N/Z)_{\rm dense}$ which is
determined by the isovector part of the nuclear mean-field.
Effects of the latter depends on the density gradients reached and
the duration of the reaction. The density range and reaction time
can be varied by varying the masses of the reaction system and the
beam energy. In doing so, to examine the effects of the HD \esym,
it is better to use reactions of the same neutron/proton ratio so
that the initial isospin asymmetry of the reaction system remains
the same and thus allows us to examine clearly effects of the
isospin fractionation. The observed effects can then be
essentially attributed to the variation of the dynamical isospin
fractionation due to the changing density and reaction time. In
the following, we compare head-on reactions of
$^{48}$Ca+$^{48}$Ca, $^{124}$Sn+$^{124}$Sn and
$^{197}$Au+$^{197}$Au at beam energies from 0.25 to 0.6 \agev.
Their isospin asymmetries are approximately the same, more
quantitatively, 1.40, 1.48 and 1.49, respectively.
\begin{figure}[h]
\centering
\includegraphics[width=\columnwidth]{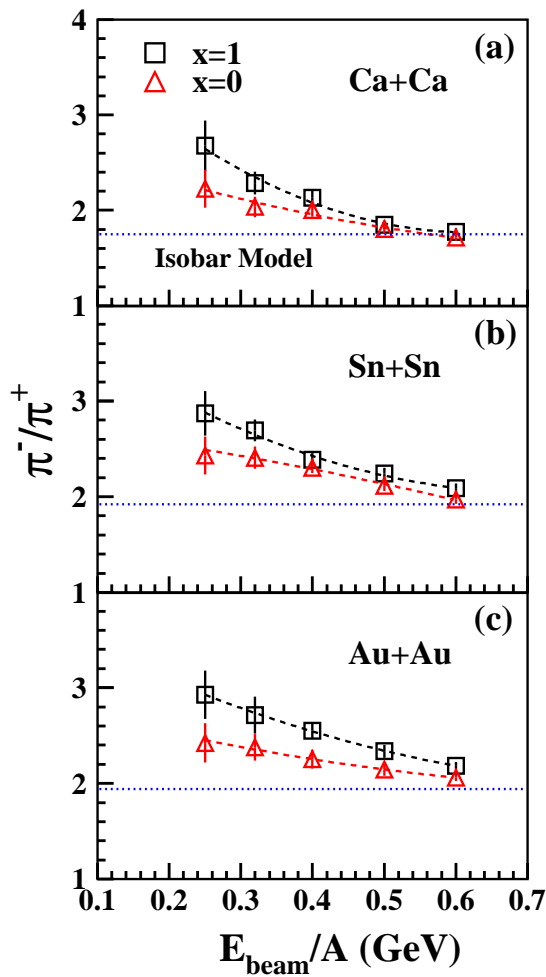}
\caption{(Color online) Excitation function of the $\pi^-$/$\pi^+$
ratio using the MDI interaction with $x=1$ (open square) and 0 (open
triangle) for the head-on collisions of $^{48}$Ca+$^{48}$Ca,
$^{124}$Sn +$^{124}$Sn and $^{197}$Au+$^{197}$Au,
respectively.}\label{fig2}
\end{figure}

\begin{figure}[h]
\centering
\includegraphics[width=\columnwidth]{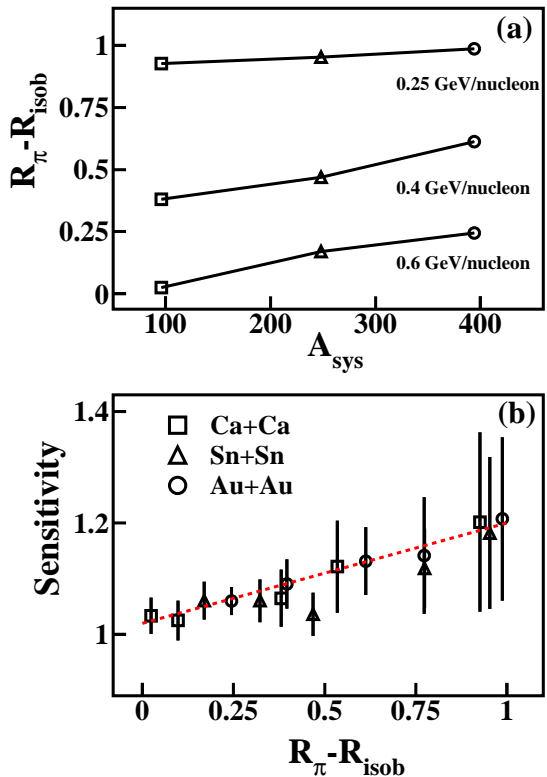}
\caption{(Color online) (a) the system size and beam energy
dependence of the degree of isospin fractionation denoted by
$R_{\pi}-R_{\rm isob}$ (see text). (b) correlation between the
sensitivity of the \rpi ratio to the \esym and the degree of isospin
fractionation. The dashed line is for guiding the eyes.}\label{fig3}
\end{figure}

Shown in Figure \ref{fig2} are the excitation functions of the
\rpi ratio with the \esym of $x=1$ and $x=0$, respectively. The
error bars presented in the plots are of only statistical origin.
The \rpi ratio increases with decreasing the beam energy and
exceeds the isobar model prediction indicated by the dashed line
in the plots. This trend is again consistent with the FOPI data in
Au+Au reactions and other model calculations \cite{WRE07}. In the
energy regime considered here, pions are mainly produced through
the $\Delta$ isobar. If we denote $R_{\pi}\equiv \pi^-/\pi^+$,
then its difference from the $\Delta$ isobar model prediction
$R_{\rm isob}$ can be used to measure effects of the isospin
fractionation on pion production. To the first order of
approximation, the $R_{\pi}-R_{\rm isob}$ is  due to the isovector
nuclear mean-field. Effects of the latter are expected to depend
on both the space-time volume and the density gradients reached in
the reaction. Shown in the upper window of Fig. \ref{fig3} is the
mass $A_{\rm sys}$ dependence of the $R_{\pi}-R_{\rm isob}$ at
beam energies of 0.25, 0.4 and 0.6 \agev, respectively. The
results are obtained using the \esym with $x=1$. The
$R_{\pi}-R_{\rm isob}$ exhibits a nearly linear dependence on the
system mass or volume at a given beam energy. At lower beam
energies, while the maximum density reached is lower the reaction
time is longer. The net isospin fractionation effect of the
isovector potential on the \rpi ratio is thus larger. On the other
hand, at higher energies, the $R_{\pi}-R_{\rm isob}$ increases
with $A_{\rm sys}$ much faster due to both the higher density
gradients reached and the larger reaction volume available in
these reactions. To quantify the sensitivity of the \rpi ratio to
the variation of the \esym and present its systematics compactly,
we define the sensitivity as the double ratio of the \rpi obtained
with the \esym of $x=1$ over that with $x=0$. The sensitivity is
shown in the lower panel of Fig.\ref{fig3} as a function of the
$R_{\pi}-R_{\rm isob}$. It is interesting to see that at a given
value of the $R_{\pi}-R_{\rm isob}$, the sensitivity is
independent of the reactions systems considered as they all have
the same initial neutron/proton ratio N/Z. Moreover, the
sensitivity increases with the $R_{\pi}-R_{\rm isob}$ as one
expects since effects of the isovector potential is approximately
linearly proportional to the isospin asymmetry of the medium.
Overall, all indications are that the \rpi ratio is probing the
strength of the isovector potential and depends on the space-time
volume of the reaction. More quantitatively, for the reactions
considered, the sensitivity increases from about several percents
to approximately twenty percents at the maximum isospin
fractionation. It suggests that the \rpi ratio in heavy systems at
relatively low beam energies is preferred for probing the HD
behavior of the \esym.

\begin{figure}[h]
\centering
\includegraphics[width=\columnwidth]{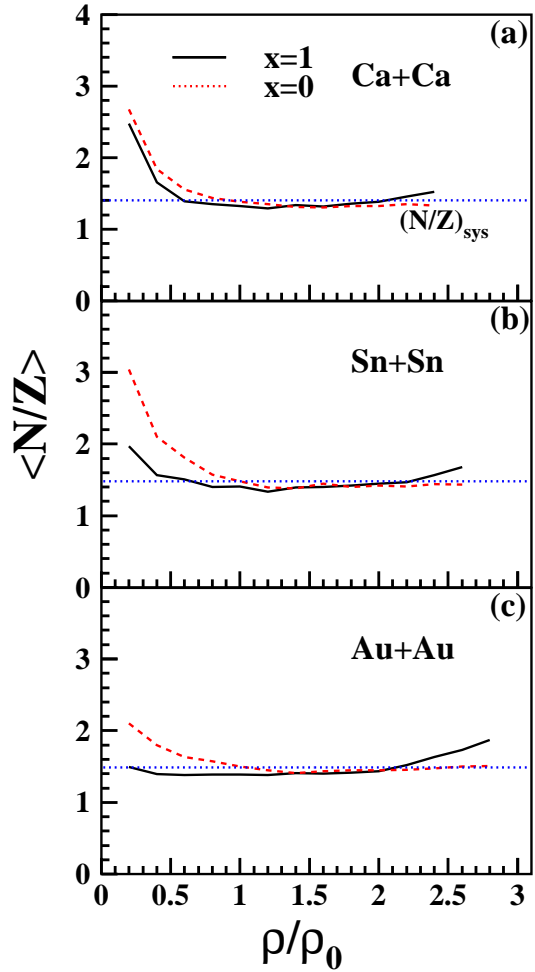}
\caption{(Color online) The evolution of the \rnz  as a function of
the local density at the most compressed stage in
$^{48}$Ca+$^{48}$Ca, $^{124}$Sn +$^{124}$Sn and
$^{197}$Au+$^{197}$Au collisions at 0.4 \agev with $x=1$ (solid
lines) and 0 (dashed lines), respectively. }\label{fig4}
\end{figure}

Having established that the \rpi ratio depends on the isovector
mean-field through the isospin fractionation mechanism, we now
examine more closely how the latter works and its dependence on
the \esym. To quantify the isospin asymmetry of an excited system
during heavy-ion reactions, we define $n_{\rm like}\equiv
n+\frac{2}{3}\Delta^0+\frac{1}{3}\Delta^+ +\Delta^-, p_{\rm
like}\equiv p+\frac{2}{3}\Delta^++\frac{1}{3}\Delta^0
+\Delta^{++}$. As a typical example, the average ratio
$<N/Z>\equiv n_{\rm like}/p_{\rm like}$ at the instant of 8, 10
and 12 fm/c when the maximum compression is reached in the
reaction of Ca+Ca, Sn+Sn and Au+Au at 0.4 \agev, respectively, is
shown as a function of the local density in Fig.\ \ref{fig4}. It
is seen that the maximum density reached is the highest for the
Au+Au reaction as one expects. To understand the results more
easily, we recall that for two regions in thermal-chemical
equilibrium at densities $\rho_1$ and $\rho_2$, and isospin
asymmetries $\delta_1$ and $\delta_2$, one has approximately
\cite{Shi00} $\delta_1\cdot E_{\rm sym}(\rho_1)=\delta_2\cdot
E_{\rm sym}(\rho_2)$. While the reaction system may not be in
complete thermal-chemical equilibrium, the above relation and the
\esym shown in Fig.\ \ref{fig1} help us understand easily the
observed correlation between the $<N/Z>$ and the local density.
Comparing the $<N/Z>$ obtained with $x=1$ and $x=0$, it is seen
that the supra-saturation density region is more neutron-rich with
$x=1$ as one expects based on the \esym shown in figure
\ref{fig1}. On the contrary, the subsaturation density region is
more neutron-rich with $x=0$. The effect from both the sub- and
supra-saturation density behavior of \esym will compete and
contribute to the final observable. As shown in figure \ref{fig2},
the fact that the \rpi ratio is higher with $x=1$ indicates that
it indeed reflects the $<N/Z>$ of the supra-saturation density
region, otherwise the final \rpi ratio would be higher with $x=0$
if it was mainly determined by the \esym at subsaturation
densities. On the other hand, it is seen in figure \ref{fig2} that
the \rpi ratios calculated with $x=0$ are also above the
predictions of the isobar model at low beam energies although the
$<N/Z>$ at supra-saturation densities is slightly below the N/Z
value of the reaction system as shown in figure \ref{fig4}. This
might indicate significant changes in the local N/Z ratios during
the time evolution of the reaction since the results in figure
\ref{fig4} represent only the density dependence of the $<N/Z>$ at
the respective instants of the maximum compression, not for the
entire duration of the reaction.

\begin{figure}[h]
\centering\includegraphics[width=\columnwidth]{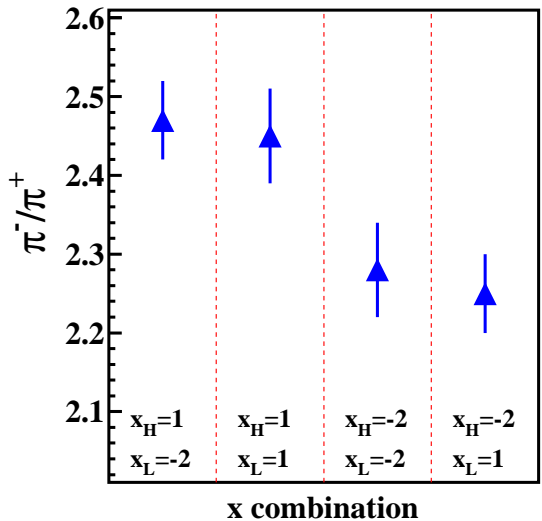}
\caption{(Color online) The $\pi^-/\pi^+$ ratio with varying \esym
at sub- ($x_L$) and supra-saturation ($x_H$) densities for the
head-on collisions of Au+Au at 400 MeV/nucleon.} \label{xlxh}
\end{figure}

Since pions always subject to strong final state interactions the
$\pi^-/\pi^+$ ratio may also be affected by the \esym at
sub-saturation densities. To quantitatively evaluate effects of
the low density \esym, let's define $x_L (x_H)$ as the $x$
parameter at sub (supra)-saturation densities, effects of the
\esym at different densities can then be revealed by comparing
calculations using different $x_L (x_H)$ parameters. As an
illustration, shown in figure \ref{xlxh} is the $\pi^-/\pi^+$
ratio at the chemical freeze-out in the Au+Au collisions at 0.4
\agev with different $x_L$ and $x_H$ combinations. Comparing the
middle two panels, it is seen that the saturated value of the
$(\pi^-/\pi^+)$ ratio increases by about 7\% when the $x$
parameter is changed from $-2$ (stiff) to $1$ (soft) at all
densities. Interestingly, if we fix the $x_H$ either at $-2$ or
$1$ but vary the $x_L$ between $1$ and $-2$, the $\pi^-/\pi^+$
ratio changes by less than 2\%. On the contrary, if the $x_L$ is
fixed, the variation of the $x_H$ leads to about a 7\% change
similar to the effect observed when the $x$ parameter is varied at
all densities also between $-2$ and $1$. Thus, again, the
$\pi^-/\pi^+$ ratio really reflects mainly the \esym at
supra-saturation densities with only very minor influences from
the \esym at sub-saturation densities.

\begin{figure}[h]
\centering
\includegraphics[width=\columnwidth]{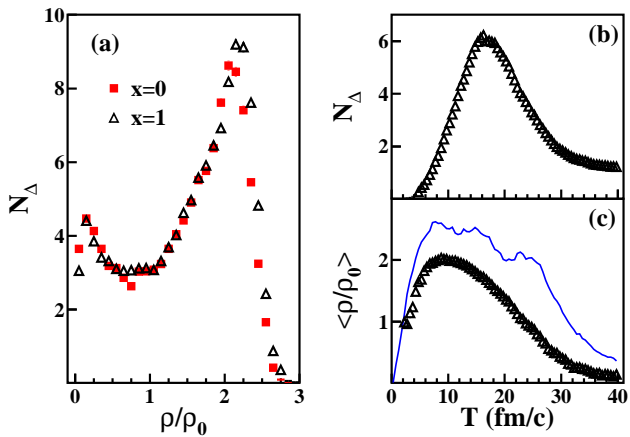}
\caption{(Color online) The distribution of $\Delta$s as a function
of density (a) and time (b), (c) the mean local density of $\Delta$s
as a function of time. The curve in (c) presents the baryon density
in the central region as a function of time.} \label{fig6}
\end{figure}

To further investigate the reliability of the \rpi ratio in
probing the HD \esym, we now turn to the density profile and time
evolution of the produced $\Delta$ resonances through both
$N+N\rightarrow N+\Delta$ and $\pi+N\rightarrow \Delta$ processes.
Shown in Figure \ref{fig6}(a) is the total multiplicity
$N_{\Delta}$ of primordial $\Delta$ resonances of all charge
states as a function of the local baryon density at which the
$\Delta$ resonances are produced during the entire reaction of
Au+Au at 400 MeV/nucleon with $x=0$ (solid square) and $x=1$ (open
triangle). It is seen that the $N_{\Delta}$ has a major peak
around $2\rho_0$ and a minor bump around $0.1\rho_0$. The latter
has contributions from both final state $\pi+N\rightarrow \Delta$
resonances and the first chance $N+N\rightarrow N+\Delta$
collisions when the surfaces of the two colliding nuclei just
start overlapping. Interestingly, the majority of $\Delta$
resonances are produced in the high density region. Moreover, the
$N_{\Delta}$ is appreciably higher with the soft \esym ($x=1$)
than that with the harder \esym ($x=0$) in the high density
region. This is consistent with the total pion multiplicity
studied in ref.\ \cite{ZGX08}. We explore next the dynamics of
resonance production. As an example, figure \ref{fig6} (b) shows
the $N_{\Delta}$ averaged over the entire reaction volume as a
function of time for the Au+Au reaction at 400 MeV/nucleon with
$x=1$. Figure \ref{fig6} (c) depicts the time evolution of the
mean local baryon density where the $\Delta$ resonances are
produced. As a reference the average baryon density in the most
central cell of 1 $fm^3$ of the colliding system is also shown
using the solid curve. By comparing the results shown in Fig.
\ref{fig6} (b) and (c), one can see clearly that the density where
the $\Delta$s are produced correlates closely to the central
baryon density in the colliding system. More quantitatively, the
$N_{\Delta}$ reaches its maximum at about 20 fm/c when the
colliding system is still in the compressed phase. Indeed, most of
the $\Delta$ resonances are produced in the supra-saturation
density region. Thus, the final \rpi ratio is more sensitive to
the supra-saturation density \esym instead of the subsaturation
one. All these demonstrate again that the \rpi ratio carries
effectively useful information about the HD \esym. Nevertheless,
it is worth noting that at the very beginning of the collision
(about 5fm/c) when the neutron skins of the projectile and the
target penetrate each other, the pion-like ratio
$(\pi^-/\pi^+)_{\rm like}$ , defined as
$(\pi^-+\Delta^-+\frac{1}{3}\Delta^0)/(\pi^++\Delta^{++}+\frac{1}{3}\Delta^+)$
which naturally becomes the $\pi^-/\pi^+$ ratio at the chemical
freeze-out, does reaches a peak higher than 3 due to the more
abundant nn collisions leading to more \pmi productions
\cite{LiBA05}, but the actual yield of pions there is very small.
Some of these pions may escape earlier along the transverse
direction, they may thus affect significantly the high energy tail
of the pion squeezed-out. However, they have only a very small
effect on the final pion ratio integrated over all space-time. It
is also necessary to mention that the study presented here uses
the free-space NN inelastic cross sections and no mean-field
effect for pions is taken into account. We notice that in-medium
effects on the NN inelastic collisions \cite {{LAR01},{LAR03}} and
pions \cite{JXU09} may affect the pion ratio and they deserve
further studies.

In summary, we carried out simulations of the head-on collisions of
$^{48}$Ca+$^{48}$Ca,$^{124}$Sn +$^{124}$Sn and $^{197}$Au+$^{197}$Au
at beam energies from 0.25  to 0.6 \agev using the IBUU04 transport
model. The $\pi^-$/$\pi^+$ ratio increases with increasing the
system size or decreasing the beam energies, indicating the
isovector mean-field is at work. A comprehensive analysis of the
isospin fractionation in these reactions having approximately the
same neutron/proton ratio but different masses, the time evolution
and the spatial distributions of the $\Delta$s demonstrate clearly
that the \rpi is an effective probe of the HD behavior of the
nuclear symmetry energy.

This work was supported in part by the National Natural Science
Foundation of China under grants 10975097, 10675148, 10635080,
10575071 and 10675082, MOE of China under project NCET-05-0392,
Shanghai Rising-Star Program under Grant No. 06QA14024, the SRF for
ROCS, SEM of China, and the National Basic Research Program of China
(973 Program) under Contract No. 2007CB815004, the US National
Science Foundation Awards PHY-0652548 and PHY-0757839, the Research
Corporation under Award No. 7123 and the Texas Coordinating Board of
Higher Education Award No. 003565-0004-2007.

\end{document}